\documentclass{WileyMSP-template}
\usepackage{amssymb,amsmath,amsthm}
\usepackage{graphicx}
\usepackage{xcolor}
\usepackage{bm,bbm}
\usepackage[bookmarks=false]{hyperref}
\hypersetup{colorlinks=true,citecolor=blue,linkcolor=red,urlcolor=blue,pdfstartview=FitH,bookmarksopen=true}
\usepackage{comment}
\usepackage{mathtools}
\usepackage{enumitem}

\usepackage{txfonts}
\usepackage[linesnumbered,ruled,vlined]{algorithm2e}
\usepackage{algpseudocode}
\usepackage{contour}
\usepackage{geometry}
\newcommand{\ket}[1]{\text{$ | #1 \rangle $}}
\newcommand{\bra}[1]{\text{$ \langle #1 | $}}

\newcommand{\tr}{\mathrm{Tr}}

\newcommand{\cD}{\mathcal{D}}

\newcommand{\rP}{\mathrm{P}}

\contourlength{0.1pt}

\newcommand{\cH}{\mathcal{H}}

\contourlength{0.1pt}

\newtheoremstyle{note}
  {\topsep/2}               
  {\topsep/2}            	  
  {}                        
  {\parindent}              
  {\itshape}                
  {.---}                    
  {0pt}                     
  {\thmname{#1}\thmnumber{ \itshape#2}\thmnote{ (#3)}} 

\newtheorem{theorem}{Theorem}

\newtheorem{proposition}[theorem]{Proposition}
\newtheorem{observation}{Observation}

\theoremstyle{definition}

\theoremstyle{remark}

\begin{document}

\pagestyle{fancy}
\rhead{\includegraphics[width=2.5cm]{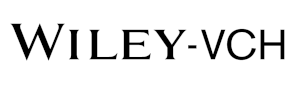}}

\title{Verification of entangled states under noisy measurements}

\maketitle

\author{Lan Zhang}
\author{Yinfei Li}
\author{Ye-Chao Liu}
\author{Jiangwei Shang*}
\begin{affiliations}
L. Zhang, Y. Li, Dr. Y.-C. Liu, Prof. J. Shang\\
Key Laboratory of Advanced Optoelectronic Quantum Architecture and Measurement, School of Physics\\
Beijing Institute of Technology\\
Beijing 100081, China\\
Email Address: jiangwei.shang@bit.edu.cn

Dr. Y.-C. Liu\\
Zuse-Institut Berlin\\
Takustra{\ss}e 7, 14195 Berlin, Germany

Dr. Y.-C. Liu\\
Naturwissenschaftlich-Technische Fakult{\"a}t\\
Universit{\"a}t Siegen\\
Walter-Flex-Stra{\ss}e 3, 57068 Siegen, Germany

Prof. J. Shang\\
State Key Laboratory of Surface Physics and Department of Physics\\
Fudan University\\
Shanghai 200433, China

\end{affiliations}
%

\keywords{quantum state verification, noise mitigation, quantum protocols}

\begin{abstract}
Entanglement plays an indispensable role in numerous quantum information and quantum computation tasks, underscoring the need for efficiently verifying entangled states. In recent years, quantum state verification has received increasing attention, yet the challenge of addressing noise effects in implementing this approach remains unsolved. In this work, a systematic assessment of the performance of quantum state verification protocols is provided in the presence of measurement noise. Based on the analysis, a necessary and sufficient condition is provided to uniquely identify the target state under noisy measurements. Moreover, this work proposes a symmetric hypothesis testing verification algorithm with noisy measurements. Then, relying on $W$ states, an SDP program is demonstrated to calculate the infidelity threshold for arbitrary measurement noise. Subsequently, using a noisy nonadaptive verification strategy of GHZ and stabilizer states, the noise effects on the verification efficiency are analytically illustrated. From both analytical and numerical perspectives, this work demonstrates that the noisy verification protocol exhibits a negative quadratic relationship between the sample complexity and the infidelity. Our method can be easily applied to real experimental settings, thereby demonstrating its promising prospects. 
\end{abstract}

\section{Introduction}\label{Sec:Intro}
The concept of quantum entanglement takes a central position in the field of quantum physics, representing a fundamental feature that sets it apart from its classical counterpart [1]. The existence of entanglement enables the quantum realm to address specific challenges more efficiently than classical computation [2]. The significance of entanglement is underscored by its wide contribution in numerous notable achievements within the field of quantum information theory. It plays a pivotal role in various applications, including quantum communication [3-6], quantum key distribution [7-9], quantum teleportation [10,11], superdense coding [12,13], quantum fault-tolerant computation [14-16], and quantum algorithms such as Grover's search algorithm [17] and Shor's algorithm [18]. Additionally, it also assumes a crucial role in quantum error correction [19-23], an indispensable component of quantum computation and information theory.

Given the indispensable role of entanglement in various tasks, the precise preparation of entangled quantum states has great significance. Hence, there is a compelling need to achieve high-precision verification of quantum states one has prepared. Conventional characterization methods, such as quantum state tomography (QST) [24-26], are highly resource intensive and time consuming. Alternative methods, like direct fidelity estimation [27,28],  attain an improvement in sample complexity as compared to QST, yet still exhibit standard quantum scaling with the infidelity. Recently, quantum state verification~(QSV)~[29] has gained much attention for providing a resource efficient approach. QSV relies solely on local measurements and classical communication, which renders it operationally convenient in experiments. Furthermore, it is crucial to emphasize that QSV is able to achieve the Heisenberg scaling in sample complexity, rather than the standard quantum limit. This renders it significantly more efficient as compared to other characterization methods.

Up to now, a number of efficient verification protocols for various entangled states have been developed, encompassing arbitrary bipartite pure states [30], Dicke states [31], phased Dicke states [32], hypergraph states [33], etc. Some of these verification protocols, such as those for bipartite pure states [34], Greenberger-Horne-Zeilinger (GHZ) states [35] and stabilizer states [36] have even been proven to be optimal in sample complexity. The verification of quantum states has also been investigated in the adversarial scenario, a context that plays an important role in various tasks including blind measurement-based quantum computation and quantum networks [37-39]. In addition to the  consideration for discrete-variable states, there have also been developments in continuous-variable state verification. These protocols have practical applications in quantum communication and quantum sensing [40]. By employing quantum nondemolition measurements, it becomes possible to enhance the sample complexity of verification protocols to the optimal global level. This approach also offers practical advantages, making it more favorable for implementation in actual experiments [41].

Not solely limited to quantum state characterization, the QSV techniques can also be effectively utilized for the purpose of quantum process verification. This includes the verification not only of standard quantum gates but also of more general quantum processes [42,43]. It is noteworthy that this method demonstrates superior applicability as compared to the conventional technique of random benchmarking [42].

However, in the noisy intermediate-scale quantum (NISQ) era [44], local measurements used in these verification protocols are imperfect. The impact of measurement noise on the performance of QSV remains an unresolved problem, particularly regarding protocol modifications and changes in sample complexity, which we will demonstrate below. In most QSV experiments [45-47], only Type I or II errors are considered independently, while the presence of measurement noise has been implicit. This ambiguity could lead to issues in the statistical analysis of sample complexity, which is crucial within the QSV framework. In Ref.[48] it assumes that QSV under noisy measurements will incur type I errors, meaning that there is a probability to reject the target state. Additionally, it provides a semidefinite programming (SDP) approach to account for the possibility of Type I error. However, how the type I error rate is influenced and to which extent the sample complexity is affected by measurement noise remain unclear.

In this work, we first conduct a systematic analysis of the impact of measurement noise on verification protocols in \textbf{Section 2}. Subsequently, a necessary and sufficient condition is provided for a noisy strategy to uniquely identify the target state.
In Section 3, we focus on symmetric testing when considering sample complexity, wherein a negative quadratic relationship between sample complexity and infidelity in this particular scenario is demonstrated. Next, $W$ state is considered as an example of a general readout noise model under which the aforementioned conditions cannot be met, then in Section 4 an SDP approach is proposed to determine the infidelity threshold at which the noisy strategy remains feasible. Then, starting with nonadaptive verification protocols for GHZ and stabilizer states, we specifically elucidate how measurement noise influences the efficiency of the verification protocols in Section 5. Finally, we conclude in Section 6.

\section{Basic Framework}\label{Sec:Framework}
The fundamental objective of QSV is to establish a well-designed protocol for verifying whether an unknown quantum state $\rho$ is the target state $\ket{\psi}$ or some bad states with the condition that $\bra{\psi}\rho\ket{\psi}\leq 1-\epsilon$~[29]. In a basic verification protocol, we measure copies of $\rho$ with an ensemble of two-outcome local projective measurements $\{\Omega_i, \mathbbm{1}-\Omega_i
\}$ equipped with a probability distribution $\{p_i\}$. The corresponding verification strategy is denoted as $\Omega = \sum_i p_i \Omega_i$. In each measurement, if $\Omega_i$ occurs the state ``passes'', otherwise the verification ``fails''. If $\rho$ is the target state $\ket{\psi}$, we require that it always passes, that is $\forall i$, $\Omega_i\ket{\psi}=\ket{\psi}$. If the copies pass consecutively many times, then the verification protocol ascertains that $\rho$ is the target state with a certain confidence level. 

In the worst-case scenario, the maximal probability that a bad state will pass the strategy $\Omega$ is~[29]
\begin{equation}
    \max _{\bra{\psi}\rho\ket{\psi}\leq 1-\epsilon} \tr(\Omega \rho) = 1-\nu(\Omega) \epsilon\,,
\end{equation}
where $\nu(\Omega)=1-\lambda_{1}$ is the spectral gap of strategy $\Omega$ between the largest eigenvalue $1$ and the second-largest eigenvalue $\lambda_{1}$. The worst-case probability for a bad state to pass the strategy $N$ times is $\left[1-\nu(\Omega) \epsilon\right]^N$.
Thus, to achieve a confidence level $1-\delta$, the sample complexity $N$ satisfies
\begin{equation}
N \geq \frac{\ln \delta^{-1}}{\ln ([1-v(\Omega) \epsilon]^{-1})}\approx \frac{1}{v(\Omega)} \epsilon^{-1} \ln \delta^{-1} \,.
\end{equation}

In standard QSV protocols, we ignore the presence of measurement noise, which is undoubtedly inconsistent with realistic scenarios. Here by the inclusion of measurement noise, we investigate when it is still possible to distinguish the null hypothesis $\cH_0$ that $\rho$ is the target state $\ket{\psi}$ from the alternative hypothesis $\cH_1$ that $\rho$ is a bad state.
Specifically, we consider the \emph{distinguishable conditions}:
\begin{enumerate}\label{distinguishable}
  \renewcommand{\theenumi}{(\alph{enumi})}
  \item $\Omega\ket{\psi}=\lambda_{0}\ket{\psi}$\,, with $\lambda_{0}\leq1$\,,
  \item $\nu(\Omega)>0$\,,
\end{enumerate}
where $\lambda_{0}$ is the largest eigenvalue of $\Omega$, $\nu(\Omega)=\lambda_0-\lambda_{1}$ is the spectral gap between $\lambda_{0}$ and the second-largest eigenvalue $\lambda_{1}$. Note that a general noisy strategy $\Omega$ may not satisfy the distinguishable conditions. For instance, it is possible that the target state is no longer an eigenstate of $\Omega$, or there exists degeneracy such that $\lambda_0 = \lambda_1$.  
\begin{observation}\label{ob1}
A QSV strategy $\Omega$ can uniquely identify a given target state $\ket{\psi}$ by probability $\bra{\psi}\Omega\ket{\psi}$ if and only if it satisfies the distinguishable conditions.
\end{observation}
The proof can be found in Appendix~A.
Observation 1 implies that if a noisy strategy $\tilde{\Omega}${, which deviates slightly from the noiseless strategy $\Omega$ due to noisy measurements,} continues to satisfy the distinguishable conditions despite the presence of noise, it remains capable of uniquely identifying the target state $\ket{\psi}$ because there must be $\tr\bigl(\tilde{\Omega}\rho\bigr)<\tr\bigl(\tilde{\Omega}\ket{\psi}\bra{\psi}\bigr)$ for any other state $\rho$. Hereinafter, we denote the verification strategy affected by noisy measurements $\tilde{\Omega}$.

Here, we relax the constraint that the target state should pass the strategy with certainty. Instead, it passes the strategy with a maximal probability $\lambda_0$.
Without loss of generality, the worst-case state for a noisy strategy could be represented by a pure state 
\begin{equation}\label{eq:WC}
\ket{\psi^{\prime}}=\sqrt{1-\epsilon}\ket{\psi}+\sqrt{\epsilon}\ket{\psi^{\perp}_{1}}\,,    
\end{equation}
and the corresponding pass probability is
\begin{equation}\label{eq:WCP}
p=\bra{\psi^{\prime}}\tilde{\Omega}\ket{\psi^{\prime}}=\lambda_0-\nu(\tilde{\Omega})\epsilon\,,    
\end{equation}
where $\ket{\psi^{\perp}_{1}}$ denotes the eigenstate of the noisy strategy $\tilde{\Omega}$ with the second-largest eigenvalue, and ${\nu(\tilde{\Omega})=\lambda_0-\lambda_{1}}$ is the spectral gap; see more details in Appendix~B. Thus, the task of discriminating $\cH_0$ and $\cH_1$ turns to distinguishing the target state $\ket{\psi}$ and the worst-case state $\ket{\psi^{\prime}}$. Note that only when the distinguishable conditions are satisfied can explicit forms of the worst-case state and the probability of the noisy strategy be obtained from \textbf{Equation 3} and \textbf{4}. Otherwise, one has to rely on SDP techniques to assess the probability of the worst-case state; further discussions can be found in Section 5.

\section{Sample Complexity}\label{Sec:SampleComp}
{ Here we consider a mixed model for the two types of errors by assuming a probability distribution of the source that prepares the unknown input states: with probability $q$, hypothesis $\mathcal{H}_0$ is true; or with probability ${1-q}$, hypothesis $\mathcal{H}_1$ is true.}
Considering measurement noise, the QSV framework can be modified as follows. Firstly, the verifier randomly chooses a local measurement $\tilde{\Omega}_i$ from the noisy strategy $\{p_i,\tilde{\Omega}_i\}$. After $N$ runs, the pass frequency $f$ is obtained, then compared with a threshold frequency $f^{\prime}$. If $f\ge f^{\prime}$, we accept $\cH_0$; otherwise we reject $\cH_0$ and accept $\cH_1$. Subsequently, the relationship between sample complexity and noise parameters is analyzed. Finally, we present an experimentally effective algorithm for this scenario.

Due to the condition $\forall i$, $\Omega_i\ket{\psi}=\ket{\psi}$ in standard QSV strategy $\Omega=\sum_i p_i \Omega_i$, it will never occur that we reject the target state $\ket{\psi}$. However, since the target states $\ket{\psi}$ cannot pass the noisy strategy $\tilde{\Omega}$ with certainty, type I error will occur, whereby the prepared target state $\ket{\psi}$ is rejected because the observed statistical frequency $f$ in the experiment is less than a predetermined threshold $f^{\prime}$. Vice versa, type II error will occur if we accept hypothesis $\cH_0$ but it is $\cH_1$ in fact; see \textbf{Table 1}.

\begin{table}[t]
\caption{Decision results of standard QSV versus noisy QSV, where `\checkmark' denotes that the event is possible, while `-' means that the event never happens. {The notation `$A | B$' represent that the verification protocol assumes hypothesis $A$ is true, while in reality, hypothesis $B$ is true.
}}\label{tab1}
\centering
\begin{tabular}{lcc}
\hline
\hline 
Decision results & Standard QSV & Noisy QSV  \\
\hline 
$\cH_0|\cH_0$ (Correct) & \checkmark & \checkmark  \\
\hline 
$\cH_1|\cH_0$ (Type I error) & - & \checkmark  \\
\hline 
$\cH_0|\cH_1$ (Type II error) & \checkmark & \checkmark  \\
\hline 
$\cH_1|\cH_1$ (Correct) & \checkmark & \checkmark  \\
\hline 
\hline
\end{tabular}
\end{table}

In order to provide an overview of the two types of errors, a simulation result is shown in \textbf{Figure 1}. This frequency distribution illustrates the verification strategy for a target state in the worst-case scenario. Depending on the threshold frequency $f^{\prime}$, the left gray shadow is type I error, meaning that we reject the target state; and the right pink shadow is type II error. If we aim for our conclusion to be sufficiently precise, it is required that both types of errors should be less than a certain confidence level.

Type I or II error follows a binomial cumulative distribution function depending on the threshold frequency $f^{\prime}$. Given a binomial random variable $X$ with success probability $p$ and the number of measurements $N$, the left binomial cumulative distribution is the probability that $X$ is less than a given value $k$, such that
\begin{equation}
\begin{aligned}
        F^{\leftarrow}(k ; N, p)&=\operatorname{Pr}(X \leq k) = \sum_{i=0}^{\lfloor k\rfloor}b_{N,i}(p)\,,
\end{aligned}
\end{equation}
where $\lfloor k\rfloor$ is the greatest integer less than or equal to $k$ and $b_{N,i}(p)=\left(\begin{array}{l}
N \\
i
\end{array}\right) p^{i}(1-p)^{N-i}$ denotes the binomial coefficient. Similarly, the right binomial cumulative distribution refers to the probability that $X$ is larger than  $k$, such that
\begin{equation}
\begin{aligned}
        F^{\rightarrow}(k ; N, p)&=\operatorname{Pr}(X \geq k) = \sum_{i=\lceil k\rceil}^{N}b_{N,i}(p)\,.
\end{aligned}
\end{equation}

\begin{figure}[t]
\includegraphics[width=.55\columnwidth]{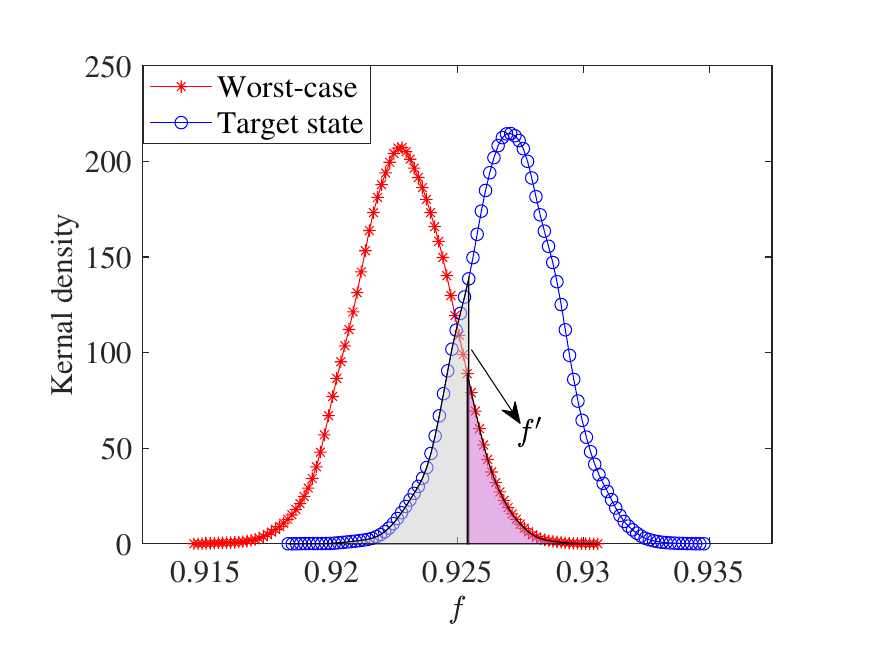}
    \centering
    \caption{Illustration of the noisy QSV with two types of errors. We consider the verification of a five-qubit stabilizer state under a noise model given in Section 5, where ${\lambda_0\sim0.9271}$ and ${\nu(\tilde{\Omega})\sim0.4341}$. The infidelity is chosen as ${\epsilon=0.01}$. The horizontal axis $f$ is the frequency of the number of passed measurements among ${N = 20000}$ measurements. The vertical axis is the kernel density function computed from $10000$ experiments. We plot the kernel density distribution for both the target state and the worst-case state. The type I and II errors correspond to the areas shaded in gray and pink respectively, which are dependent on the threshold frequency $f^{\prime}$ that we set, shown as the black vertical line.}\label{frequency}
\end{figure}

Hence, type I and II errors are given by $F^{\leftarrow}\bigl(f^{\prime}N ; N, \lambda_0\bigr)$ and $F^{\rightarrow}\bigl(f^{\prime}N ; N, \lambda_0-\nu(\tilde{\Omega})\epsilon\bigr)$ respectively. 
If we take symmetric testing, that is to consider both type I and II errors simultaneously, the average error rate is
\begin{equation}
\rP_\text{ave}=qF^{\leftarrow}\bigl(f^{\prime}N ; N, \lambda_0\bigr)+(1-q)F^{\rightarrow}\bigl(f^{\prime}N ; N, \lambda_0-\nu(\tilde{\Omega})\epsilon\bigr)\,,
\end{equation}
with ${q\in\left[0,1\right]}$.
Assume that the probability for either of the two types of errors to occur is the same, the average error rate is
\begin{equation}\label{Psym}
\rP_\text{sym}=\frac{F^{\leftarrow}\bigl(f^{\prime}N ; N, \lambda_0\bigr)+F^{\rightarrow}\bigl(f^{\prime}N ; N, \lambda_0-\nu(\tilde{\Omega})\epsilon\bigr)}{2}\,.
\end{equation}
According to Figure 1, the optimal choice of $f^{\prime}$ should be the intersection point of the two binomial distributions, as $\rP_\text{sym}$ is the sum of two shaded areas and its minimal value is obtained when $f^{\prime}$ is the intersection of the two binomial distributions. For simplicity, we choose 
\begin{equation}
    f^{\prime}=\frac{\lambda_0+\lambda_0-\nu(\tilde{\Omega})\epsilon}{2}
\end{equation}
as the threshold frequency to compute the error rate $\rP_\text{sym}$. 
This is reasonable as the two kernel distribution curves have similar shapes when $\nu(\tilde{\Omega})\epsilon$ is small enough as compared to $\lambda_0$.

In order to achieve the confidence level $1-\delta$, it is required that
\begin{equation}\label{eq:SC}
\rP_\text{sym}\leq\delta\,,    
\end{equation}
which concludes the sample complexity of the noisy QSV protocol. 
A numerical simulation is illustrated in \textbf{Figure 2}, which indicates that the sample complexity $N$ exhibits a nearly negative quadratic relationship with the infidelity $\epsilon$, i.e., ${N \sim \epsilon^{-2}}$ under noisy QSV. This relationship becomes more evident when considering the Chernoff bound as a binomial tail bound; see the details in Appendix~C. The theoretical results we have obtained differ from certain experimental findings [45-47], where the sample complexity $N$ is approximately ${N \sim \epsilon^{-1}}$. In these experiments, the error rate, denoted by $\rP$, is described by
\begin{equation}\label{eq:SCSIN}
\rP = e^{-N\cD(f \mid\mid 1-\nu(\Omega)\epsilon)} \leq \delta\,.
\end{equation}
Note that \textbf{Equation 11} adopts asymmetric hypothesis testing, i.e., only Type I or Type II error is considered, whereas our approach involves symmetric hypothesis testing to determine the sample complexity. Moreover, as compared to the scenario of \textbf{Equation 10}, the impact of noisy measurement in Equation 11 is modified from $1-\nu\epsilon$ to $\lambda_0 - \nu\epsilon$, indicating that the measurement noise contributes to decrease the eigenvalue $\lambda_0$ of the target state.

\begin{figure}[t]
    \centering
    \includegraphics[width=.55\columnwidth]{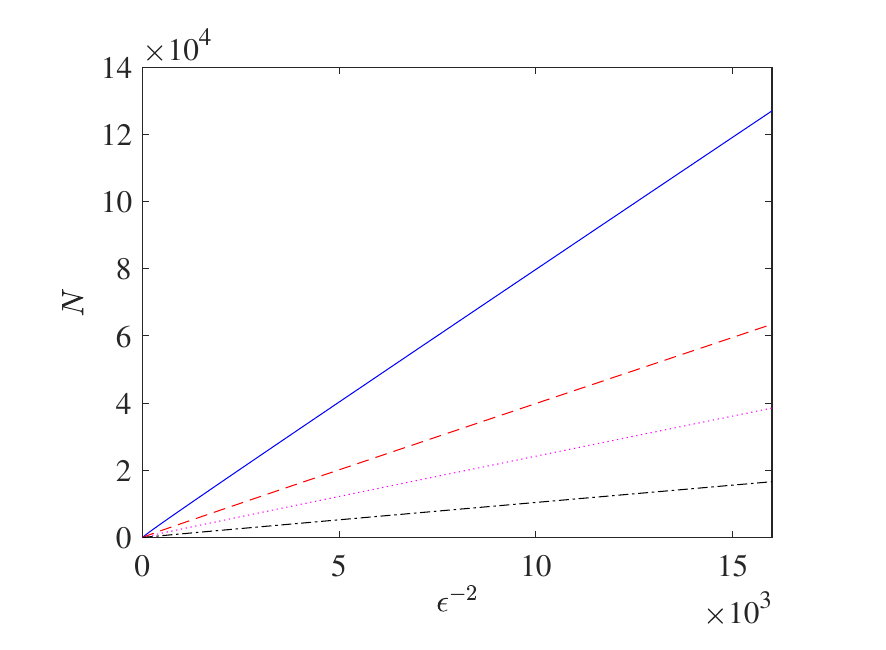}
    \caption{Noisy verification of a five-qubit stabilizer state. The simulated noisy strategy has the dominant eigenvalue ${\lambda_0\sim0.9271}$ and the second-largest eigenvalue ${\lambda_{1}\sim0.4930}$. Four curves are plotted for different confidence levels $\delta=0.01, 0.05, 0.1, 0.2$ (top to bottom).}\label{figSC}
\end{figure}

The results indicate that if we relax the completeness condition, i.e., the verification protocol is not necessarily required to accept the target state, the sample complexity immediately exhibits a linear relationship with $\epsilon^{-2}$, reaching the standard quantum limit. 
Based on the discussion, we can enhance the QSV protocol for greater precision and effectiveness in experiments, as outlined in \textbf{Algorithm 1}.

\begin{algorithm}[h]
 \SetAlgoLined 
\caption{Noisy QSV}\label{alog1}    

\KwIn {The unknown state $\rho$, the infidelity $\epsilon$, the confidence level $1-\delta$, and a noisy strategy $\tilde{\Omega}=\sum_i p_i\tilde{\Omega}_i$ satisfying the distinguishable conditions.}
\KwOut  {Determine whether $\cH_0$ or $\cH_1$ is true with confidence level $1-\delta$.}      
Calculate the dominant eigenvalue $\lambda_0$ and the spectral gap $\nu(\tilde{\Omega})$ of $\tilde{\Omega}$.\

Get the number of measurements $N$ according to Equation 10.\

Randomly employ a test $\tilde{\Omega}_i$ from $\tilde{\Omega}$ with probability $p_i$ for $N$ runs.\

Calculate the pass frequency $f$.\
    
If $f\geq f'$, accept that the unknown state $\rho$ is in $\cH_0$; otherwise, accept that the unknown state is in $\cH_1$.

\end{algorithm}

\section{General scenarios}\label{Sec:General}
In the preceding section, we have discussed the noisy verification strategy that satisfies the distinguishable conditions, whereas for general target state verification and noise models such as the coherent noise, these conditions may not be fulfilled. Nevertheless, it is still possible to distinguish the null hypothesis $\cH_0$ from the alternative hypothesis $\cH_1$ as illustrated below.

For an $n$-qubit $W$ state
\begin{equation}
\ket{W_n}=\frac{1}{\sqrt{n}}\Bigl(\ket{0\cdots01}+\ket{0\cdots10}+\cdots+\ket{1\cdots00}\Bigr)\,,
\end{equation}
there exists a nonadaptive verification strategy [31]
\begin{equation}
    \Omega_{W}=\frac{1}{2} \mathcal{Z}^{1}+\frac{1}{2 C_{n}^{2}} \sum_{i<j} \Omega_{i, j}\,,
\end{equation}
with $\mathcal{Z}^{1}=\sum_{u \in B_{n, 1}}\ket{u}\bra{u}$, $\Omega_{i, j}=\overline{\mathcal{Z}}_{i, j}^{0}(X X)_{i, j}^{+}+\overline{\mathcal{Z}}_{i, j}^{1}(\mathbbm{1} \mathbbm{1})_{i, j}$. $B_{n, 1}$ denotes the set of strings in $\{0, 1\}^n$ with Hamming weight 1, and $\overline{\mathcal{Z}}_{i, j}^{k}$ means that $k$ excitations are observed in Pauli-$Z$ measurement except for $i$ and $j$. Under the measurement noise as described in \textbf{Equation 24}, the noisy strategy $\tilde{\Omega}_{W}$ is given by
\begin{equation}
    \tilde{\Omega}_{W}=\frac{1}{2} \tilde{\mathcal{Z}}^{1}+\frac{1}{2 C_{n}^{2}} \sum_{i<j} \tilde{\Omega}_{i, j}\,,
\end{equation}
where $\tilde{\mathcal{Z}}^{1}=\sum_{u \in B_{n, 1}}\tilde{\Pi}_{u}$, with $\tilde{\Pi}_{u}$ being the noisy version of the projector $\ket{u}\bra{u}$. $\tilde{\Omega}_{i, j}$ is similar to $\tilde{\mathcal{Z}}^{1}$, except for the measurements that are on the $i$-th and $j$-th qubits.

\begin{figure}[t]
    \includegraphics[width=.55\columnwidth]{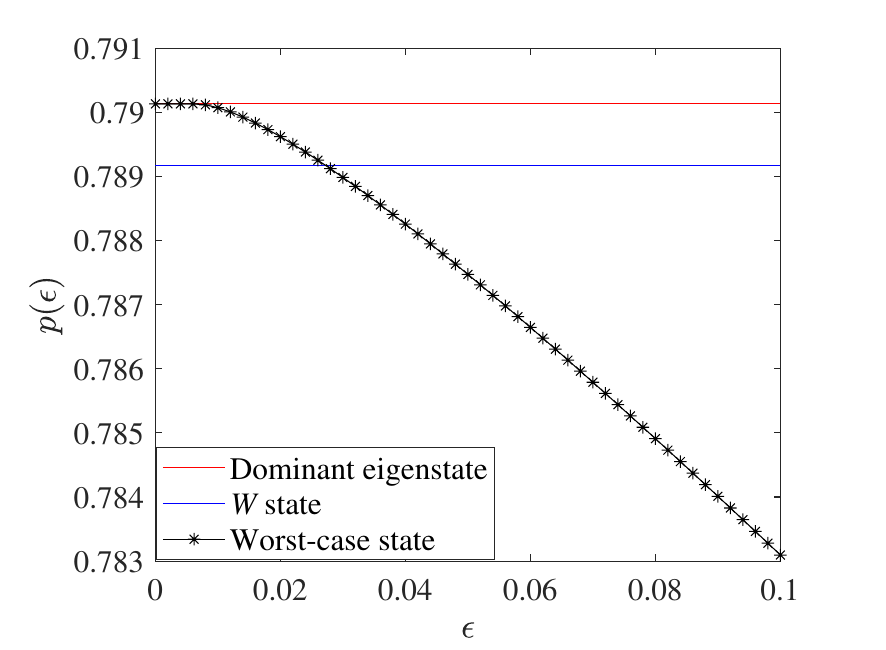}
    \centering
    \caption{Numerical results on a three-qubit $W$ state noisy verification. The vertical axis labels the probability $p(\epsilon)$ of passing noisy nonadaptive strategy $\tilde{\Omega}_{W}$ and the horizontal axis denotes the infidelity $\epsilon$. The red line is the dominant eigenvalue of noisy strategy $\tilde{\Omega}_{W}$. The blue line represents the passing probability of the noisy strategy $\tilde{\Omega}_{W}$ for the $W$ state. The black line indicates the passing probability calculated for the worst-case scenario using the SDP in Equation 17.}\label{Wsimulation}
\end{figure}

For a more practical form of the noisy projective measurement $\tilde{\Omega}_{W}$, the following readout noise model has been considered~[49]
\begin{equation}
    \begin{pmatrix}
 \tilde{\Pi}_1\\\tilde{\Pi}_2
 \\\vdots

\end{pmatrix}=\Lambda\begin{pmatrix}
 \Pi_1\\\Pi_2
 \\\vdots
\end{pmatrix}+\Delta\,,
\end{equation}
where $\Pi_i$s are ideal measurements satisfying ${\sum_i\Pi_i=\mathbbm{1}}$. {The left stochastic matrix \(\Lambda\) represent a classical transition error, which is equivalent to classically post-processing the statistics. The vector \(\Delta\) represents the residual component of certain types of quantum measurement noise, which cannot be treated as the classical transition error $\Lambda$.
} During the experiment, the ideal measurement $\Pi_i$ is replaced by the noisy measurement $\tilde{\Pi}_i$.
For example, the noisy POVM elements of a two-outcome ideal measurement $\{\Pi_1\, ,\Pi_2\}$ are
\begin{equation}
\begin{aligned}
\tilde{\Pi}_1&=(1-\eta)\Pi_1+q\Pi_2+\Delta_1\,,\\
\tilde{\Pi}_2&=\eta\Pi_1+(1-q)\Pi_2+\Delta_2\,.
\end{aligned}
\end{equation}
Note that the noisy POVM $\bigl\{\tilde{\Pi}_i\bigr\}$ is still valid, implying that both $\tilde{\Pi}_1$ and $\tilde{\Pi}_2$ are semidefinite positive operators satisfying $\tilde{\Pi}_1+\tilde{\Pi}_2=\mathbbm{1}$.

Generally $\ket{W_n}$ is not an eigenstate of the noisy strategy $\tilde{\Omega}_{W}$ if there is no specific assumption on the noisy parameters $\Lambda$ and $\Delta$. Its passing probability ${\lambda^{\prime}=\tr\Bigl(\tilde{\Omega}_{W}\ket{W_n}\bra{W_n}\Bigr)}$ is smaller than the dominant eigenvalue $\lambda_0$ of the noisy strategy $\tilde{\Omega}_{W}$, which indicates that the distinguishable conditions are not met. By using semidefinite programming, the worst-case pass probability $p(\epsilon)$ can be obtained,
\begin{equation}\label{eq:SDPP}
p(\epsilon)=\max_{\substack{\bra{\psi}\rho\ket{\psi}\leq 1-\epsilon \\ \rho \succeq 0 \\ \tr(\rho) = 1}}\tr(\rho\tilde{\Omega})\,,
\end{equation}
which depends on the specific noisy verification strategy $\tilde{\Omega}$. The infidelity threshold $\epsilon_\text{th}$ is given by $\max_\epsilon p(\epsilon) = \lambda'$. When ${\epsilon>\epsilon_\text{th}}$, there is ${p(\epsilon) < \lambda^{\prime}}$, meaning that the passing probability of the target state $\ket{\psi}$ is greater than any bad state in $\cH_1$. Thus the target state $\ket{\psi}$ could be verified using Algorithm~\ref{alog1} with ${f^{\prime}=\frac{p(\epsilon)+\lambda^{\prime}}{2}}$ and $\nu(\tilde{\Omega})=\lambda^{\prime}-\epsilon$. The sample complexity is numerically provided by Equation 10. The numerical results are depicted in \textbf{Figure 3}, where the infidelity threshold is ${\epsilon_\text{th} \approx 0.03}$. This means that when ${\epsilon > 0.03}$, the noisy strategy $\tilde{\Omega}_{W}$ is still feasible to verify $\ket{W_n}$ from the bad case. This numerical result depends on the parameter setting that $\Delta=\textbf{0}$, and $\eta, q$ are random between $[0,0.3]$. To summarize, the infidelity threshold $\epsilon_\text{th}$ can be determined by the following algorithm:

\begin{algorithm}[h]
 \SetAlgoLined 
\caption{Calculate $\epsilon_\text{th}$ and $N$} \label{alog2}    

\KwIn {The target state $\ket{\psi}$, the confidence level $1-\delta$ and a noisy strategy $\tilde{\Omega}=\sum_i p_i\tilde{\Omega}_i$}
\KwOut  {The infidelity threshold $\epsilon_\text{th}$ satisfying $p(\epsilon_\text{th})=\tr(\tilde{\Omega}\ket{\psi}\bra{\psi})$}

Solve $p(\epsilon)=\tr(\tilde{\Omega}\ket{\psi}\bra{\psi})$ in the region $\epsilon \in [0,1]$ numerically, and $p(\epsilon)$ is solved by semidefinite programming.\

Compare all the $\epsilon$ and output the largest one as $\epsilon_\text{th}$.\

Calculate the sample complexity $N$ according to Equation 10 with ${f^{\prime}=\frac{p(\epsilon)+\lambda^{\prime}}{2}}$ and $\nu(\tilde{\Omega})=\lambda^{\prime}-p(\epsilon)$.\

\end{algorithm}

Because of the monotonicity of $p(\epsilon)$, the existence of an infidelity threshold $\epsilon_\text{th}$ requires $p(1)<\lambda^{\prime}$, which means the pass probability of any state orthogonal to the target state $\ket{\psi}$ is less than $\lambda^{\prime}$. Since all states orthogonal to the target state are $\mathbbm{1}-\ket{\psi}\bra{\psi}$ and includes $2^n-1$ states, we have $\tr\bigl[(\mathbbm{1}-\ket{\psi}\bra{\psi})\tilde{\Omega}\bigr] < (2^n-1) \lambda^{\prime} $. This implies that
\begin{equation}\label{eq:trace}
    \tr\bigl(\tilde{\Omega} \bigr) < 2^n\lambda^{\prime}
\end{equation}
holds true. Note that \textbf{Equation 18} is a necessary condition for the existence of infidelity threshold $\epsilon_\text{th}$. If Equation 18 is violated, $\epsilon_\text{th}$ does not exist. However, even if Equation 18 holds, we still need to employ \textbf{Algorithm 2} to obtain the threshold $\epsilon_\text{th}$.

\section{Distinguishable conditions}\label{Sec:Dist}
The infidelity threshold $\epsilon_\text{th}$ can be determined by Algorithm 2 regardless of the measurement noise. If certain special assumptions on the measurement noise are accessible, we can analytically verify the target state without relying on the conditional infidelity threshold $\epsilon_\text{th}$ provided by Algorithm 2. To demonstrate the validity of the theoretical framework, we conduct verification experiments on five-qubit stabilizer states and GHZ states. For analytical convenience, the noise amplitudes are intentionally set along the $X$, $Y$, and $Z$ directions of the local Pauli measurements to be the same. These experiments aim to illustrate the correlation between the noise amplitude and confidence level under given conditions, notably with infidelity ${\epsilon=0.01}$ and number of measurements ${N=20000}$.

An $n$-qubit stabilizer state $\ket{\psi}$ with its stabilizer group $G=\{G_k\}$ can be verified by the following strategy~[29]
\begin{equation}\label{staveri}
\begin{aligned}
\Omega_{\mathrm{S}}&=\frac{1}{2^n-1}\sum_{G_k\in G\setminus\{\mathbbm{1}\} }E_k\,,
\end{aligned}   
\end{equation}
where $E_k=\frac{1}{2}\bigl(G_k+\mathbbm{1}\bigl)
$ corresponds to the projection onto the positive eigensubspace of the group element $G_k$. To deduce a more clearer description of the sample complexity in Equation 10, we consider a special assumption when $\eta = q$ and $\Delta=\textbf{0}$, that is
\begin{equation}\label{Eq:qubitNoise}
 \begin{pmatrix}
 \tilde{\Pi}_1\\\tilde{\Pi}_2
\end{pmatrix}=\begin{pmatrix}
 1-\eta&\eta\\\eta&1-\eta
\end{pmatrix}
\begin{pmatrix}
 \Pi_1\\\Pi_2
\end{pmatrix}.   
\end{equation}
It follows that under this condition, readout noise will only affect the eigenvalues with a noise factor, which we denote by $g$. Taking Pauli-$X$ measurement as an example with
\begin{equation}
 \begin{aligned}
    \Pi_{X+}&=\frac{\mathbbm{1}+X}{2}\,,\quad
    \Pi_{X-} =\frac{\mathbbm{1}-X}{2} \,,
\end{aligned}   
\end{equation}
the noisy measurements are
\begin{equation}
\begin{aligned}
    \tilde{\Pi}_{X+} & = \frac{1}{2}\bigl[\mathbbm{1}+(1-2\eta)X\bigr]\,,\\
    \tilde{\Pi}_{X-}& = \frac{1}{2}\bigl[\mathbbm{1}-(1-2\eta)X\bigr]\,.
\end{aligned}
\end{equation}
Compared with the ideal measurement, $\Big\{\tilde{\Pi}_{X+}, \tilde{\Pi}_{X-}\Big\}$ only differs by a noise factor $g=1-2\eta$.

For $n$-qubit Pauli measurements $\bigg\{\frac{\mathbbm{1}+\otimes_i^n P_i}{2},\frac{\mathbbm{1}-\otimes_i^n P_i}{2}\bigg\}$ with $P_i \in\{X,Y,Z\}$ acting on the $i$-th qubit, we have
\begin{equation}
\begin{aligned}
    &\frac{\mathbbm{1}+{\otimes}_i^n P_i}{2}
    =\frac{1}{2}\bigg({\otimes}_i^n \Big(\Pi_i (+)+\Pi_i (-)\Big)+{\otimes}_i^n \Big(\Pi_i (+)-\Pi_i (-)\Big)\bigg)
\end{aligned}
\end{equation}
where $\Pi_i(+)$ and $\Pi_i(-)$ are positive and negative projective measurements on the $i$-th qubit respectively.
If the readout errors affecting each qubit are uncorrelated, we have the noise model [49]
\begin{equation}\label{Eq:uncorrelated}
    \Lambda={\otimes}_i \Lambda_i\,,
\end{equation}
where $\Lambda_i$ is the readout noise affecting the two-outcome measurement on the $i$-th qubit. Moreover, if each $\Lambda_i$ is analogous to \textbf{Equation 20}, the noisy version of measurement $\frac{\mathbbm{1}+\otimes_i^n P_i}{2}$ reads
\begin{equation}\label{maltinoisy}
\begin{aligned}
&\frac{{\otimes}_i^n \Big(\tilde{\Pi}_i(+)+\tilde{\Pi}_i (-)\Big) + {\otimes}_i^n \Big(\tilde{\Pi}_i(+)-\tilde{\Pi}_i(-)\Big)}{2}
=\frac{\mathbbm{1}+\prod_i^n(1-2\eta_i){\otimes}_i^n P_i}{2}
\end{aligned}
\end{equation}
with $\Lambda_i=\begin{pmatrix}
 1-\eta_i&\eta_i\\\eta_i&1-\eta_i
\end{pmatrix}$, which still acts as a noise factor $g=\prod_i^n(1-2\eta_i)$ before the Pauli operator ${\otimes}_i^n P_i$.

Under the assumptions in Equation 20 and 24, measurement noise only introduces a noise factor $g$ which depends on the multi-qubit Pauli measurement. Then \textbf{Proposition 1} in the following is evident. Note that Equation 20 and  24 serve as sufficient conditions only for the distinguishable conditions. In Section 5 we discuss the scenario where the distinguishable conditions are not met by considering $W$ state as an example.

\begin{proposition}\label{prop1}
The verification strategy $\Omega_{\mathrm{S}}$ of the stabilizer states $\ket{\psi}$ could maintain the distinguishable conditions if the readout noise satisfies the conditions in Equation 20 and 24. The dominant eigenvalue is $\frac{1}{2^n-1}\sum_k\frac{1}{2}(g_k+1)$, where $g_k$ denotes the noise parameter of the Pauli measurement.
\end{proposition}

The proof can be found in Appendix~D.
Proposition 1 guarantees that the verification strategy in \textbf{Equation 19} retains its ability to verify the target state even in the presence of measurement noise and also provides the spectral gap $\nu(\tilde{\Omega})$ used in Algorithm 1. {For the simulation results shown in \textbf{Figure 4}, we consider the noisy verification of a five-qubit stabilizer state $\rho_{G_k}$, with its stabilizer group $\{G_k\}$ generated by $\langle XZZX\mathbbm{1}, \mathbbm{1}XZZX, X\mathbbm{1}XZZ, ZX\mathbbm{1}XZ, ZZZZZ \rangle$. For a given noise amplitude $g$, we randomly generate either the target state $\rho_{G_k}$ or the worst-case state $\rho_{G_k}^{\prime}$, as described in Equation (4). The unknown state is then verified using the noisy strategy $\tilde{\Omega}$ as defined in \textbf{Equation 25}, with a sample complexity of $N = 20000$. 

The simulation results illustrate the ratio of the correct verification cases. Additionally, the theoretical error rate $\rP_\text{sym}$ is calculated based on \textbf{Equation 10}. The entire process is detailed in \textbf{Algorithm 3} below.

} Figure 4 shows that the confidence level depends negatively with the noise amplitude, which supports Equation 10.

\begin{algorithm}[h]
 \SetAlgoLined 
\caption{Simulated experiment on noisy QSV}\label{alog3}    

\KwIn {The unknown machine randomly generates target state $\ket{\psi}$ or worst-case state $\rho$ with the infidelity $\epsilon$ and a noisy strategy $\tilde{\Omega}=\sum_i p_i\tilde{\Omega}_i$ depending on a noisy parameter $g$.}
\KwOut  {Determine whether $\cH_0$ or $\cH_1$ is true with confidence level $1-\delta$.}
Generate state $\rho$ as either the target state $\ket{\psi}$ or a worst-case noisy state with equal probabilities.\

Randomly employ a test $\tilde{\Omega}_i$ from $\tilde{\Omega}$ with probability $p_i$ for $N=20000$ runs.\

We accept or reject the quantum state $\rho$ generated in step 1 with the verification protocol in Algorithm 1. If neither type I nor type II error occurs in the verification result, we call it correct.\
    
Repeat Step 1 to Step 3 10000 times, simulated confidence level is the percentage of the correct results.\

Calculate the theoretical confidence level according to Equation 10.

\end{algorithm}
\begin{figure}[t]
    \includegraphics[width=.55\columnwidth]{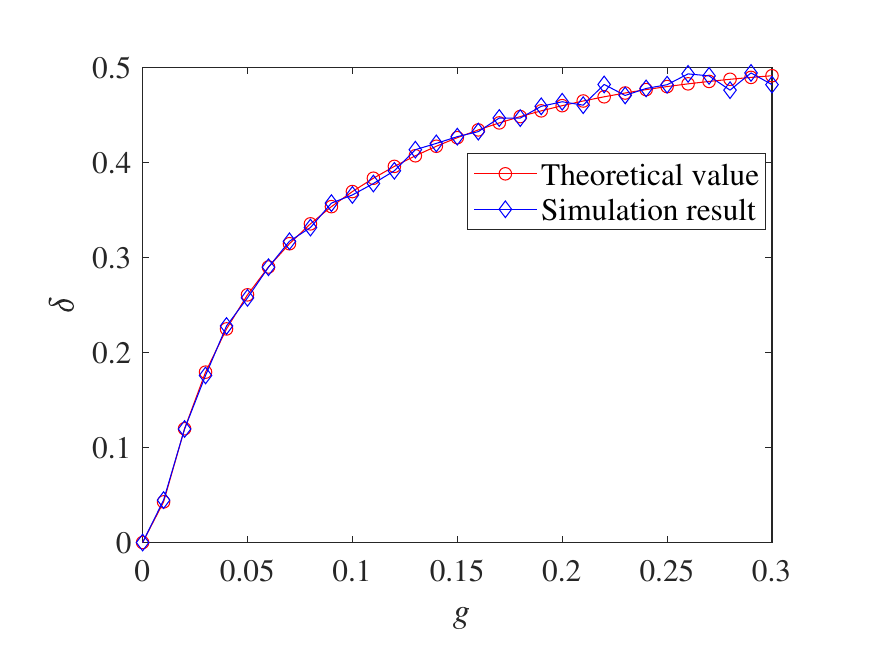}
    \centering
    \caption{Simulation and theoretical results on a five-qubit stabilizer state noisy verification. The vertical axis labels the confidence level $\delta$ and the horizontal axis $g$ denotes the noisy amplitude. For convenience, the noise amplitude is chosen to be the same in all directions. The simulation settings are of sample complexity $N=20000$, infidelity $\epsilon=0.01$ with 10000 times repetition.}\label{stafig}
\end{figure}

In another example we consider $n$-qubit GHZ states, 
\begin{equation}
    \ket{\text{GHZ}_n}=\frac{1}{\sqrt{2}}\Bigl(\ket{0}^{\otimes n}+\ket{1}^{\otimes n}\Bigr)\,.
\end{equation}
The verification strategy $\Omega_{\mathrm{I}}$ reads [35]
\begin{equation}\label{GHZveri}
    \Omega_{\mathrm{I}}:=\frac{1}{3}\left(P_{0}+\frac{1}{2^{n-2}} \sum_{\mathcal{Y}} P_{\mathcal{Y}}\right)=\frac{1}{3}\Bigl(\mathbbm{1}+2\ket{\text{GHZ}_{n}}\bra{\text{GHZ}_{n}}\Bigr)\,,
\end{equation}
where 
\begin{equation}
\begin{aligned}  
P_0 &=\bigl(\ket{0}\bra{0}\bigr)^{\otimes{n}}+\bigl(\ket{1}\bra{1}\bigr)^{\otimes{n}}\,,\\
P_{\mathcal{Y}} &=\frac{1}{2}\left(\mathbbm{1}+(-1)^{t} \prod_{k \in \mathcal{Y}} Y_{k} \prod_{k^{\prime} \in \bar{\mathcal{Y}}} X_{k^{\prime}}\right).
\end{aligned}
\end{equation} 
The spectral gap is ${\nu\bigl(\Omega_{\mathrm{I}}\bigr)=2/3}$. Based on the aforementioned discussion of the readout noise, the noisy strategy $\tilde{\Omega}_{\mathrm{I}}$ can be expressed as
\begin{equation}
\begin{aligned}
&\tilde{\Omega}_{\mathrm{I}}=\frac{1}{3}\left(\tilde{P}_{0}+\frac{1}{2^{n-2}} \sum_{\mathcal{Y}} \tilde{P}_{\mathcal{Y}}\right)=\frac{1}{3}\left[\tilde{P}_{0}+\frac{1}{2^{n-2}} \sum_{\mathcal{Y}} \frac{1}{2}\left(\mathbbm{1}+(-1)^{t}g_{\mathcal{Y}} \prod_{k \in \mathcal{Y}} Y_{k} \prod_{k^{\prime} \in \bar{\mathcal{Y}}} X_{k^{\prime}}\right)\right],
\end{aligned}
\end{equation}
where $\tilde{P}_{0}$ and $\tilde{P}_{\mathcal{Y}}$ are the noisy versions of the tests ${P}_{0}$ and ${P}_{\mathcal{Y}}$, respectively. The expression of $\tilde{P}_{0}$ is displayed in \textbf{Equation 52} in Appendix~E. Similarly, if the readout noise can maintain certain special conditions as in Equation 20 and 24, the noisy strategy $\tilde{\Omega}_{\mathrm{I}}$ could satisfy the distinguishable conditions, as outlined in the following proposition.
\begin{proposition}\label{prop2}
    The verification strategy $\Omega_{\mathrm{I}}$ for GHZ state satisfies the distinguishable conditions for readout noise stated in Equation 20,  24, and  \textbf{53}.
\end{proposition}
\begin{figure}[t]
    \includegraphics[width=.55\columnwidth]{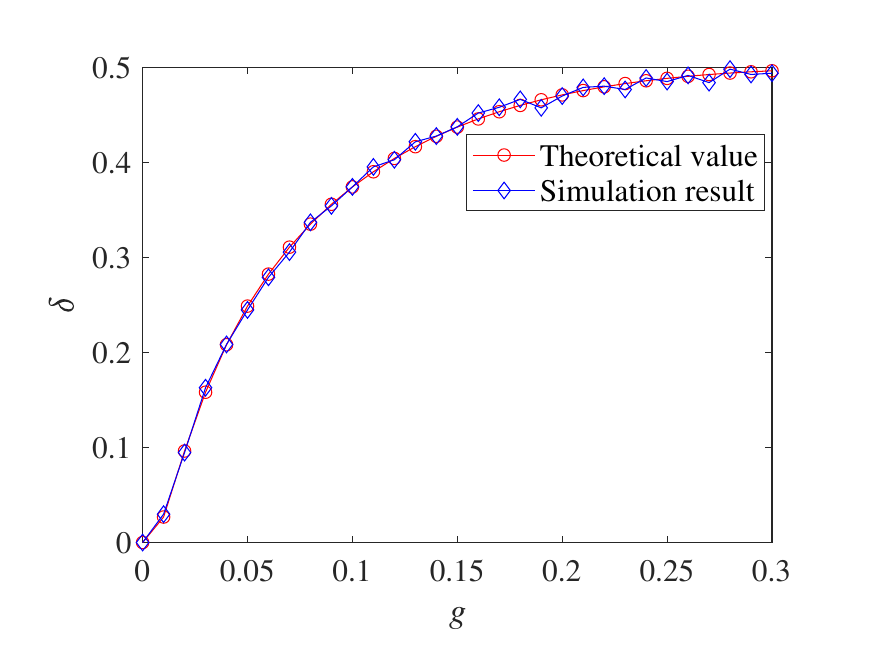}
    \centering
    \caption{Simulation and theoretical results on a five-qubit GHZ state noisy verification. The vertical axis labels the confidence level $\delta$ and the horizontal axis $g$ denotes the noisy amplitude. For convenience, the noise amplitude is chosen to be the same in all directions. The simulation experiment settings are of sample complexity $N=20000$, infidelity $\epsilon=0.01$ with 10000 times repetition.}\label{GHZfig}
\end{figure}

The proof can be found in Appendix~E.
Similar to Proposition 1, Proposition~\ref{prop2} demonstrates that with noisy measurements, strategy $\Omega_{\mathrm{I}}$ is still able to uniquely distinguish GHZ states. Similar to what is shown in Figure 4, \textbf{Figure 5} demonstrates that with ${N=20000}$ measurements, high
precision verification can be achieved when the noise amplitude $g_k$ is relatively low. As the noise amplitude $g_k$ increases, we can maintain a given confidence level by increasing the number of measurements. This also illustrates the correctness of the theoretical framework.

As can be seen, even if the sample complexity degrades to the standard quantum scaling, it still grows polynomially with the number of measurements $N$, implying that our protocol still enables efficient verification of entangled states through noisy measurements. Note that Propositions 1 and 2 rely on the distinguishable conditions. For those scenarios without the distinguishable conditions, we will demonstrate in the next section that a noisy strategy may still be capable of distinguishing the target state $\ket{\psi}$ from bad states in $\cH_1$, which can be ascertained by semidefinite programming.


\section{Conclusion}\label{Sec:Con}
Efficient verification of entangled states with imperfect measurement apparatus is of practical interest for various quantum information processing tasks. In this work, we investigated the QSV protocols under noisy measurements. When the distinguishable conditions are met, the noisy QSV strategy is still faithful to verify the target state. However, due to the measurement noise, both type I and type II errors can happen in the hypothesis testing framework as described in Algorithm 1.
Our results show that the sample complexity $N$ exhibits a negative quadratic relation with the infidelity $\epsilon$ for symmetric hypothesis testing. Moreover, for arbitrary measurement noise that violates the distinguishability conditions, a general numerical approach is provided to identify the infidelity threshold, below which the target state can still be verified. Using $W$ states, the sample complexity corresponding to the confidence level $1-\delta$ is also determined. Additionally, we provide a theoretical analytical solution based on specific assumptions about the measurement noise, supplemented with numerical experiments on stabilizer states and GHZ states. Our work illustrates the potential for efficient verification of various entangled states of interest using existing verification strategies, avoiding the need for more complex experimental settings. Furthermore, our protocols have the potential to support robust noisy verification in the future.

\medskip
\textbf{Supporting Information} \par 
The data that support the findings of this study are available from the corresponding author upon reasonable request.

\medskip
\textbf{Acknowledgements} \par 
This work was supported by the National Natural Science Foundation of China (Grants No.~12175014 and No.~92265115) and the National Key R\&D Program of China (Grant No.~2022YFA1404900).
Y.-C. Liu is also supported by the Deutsche Forschungsgemeinschaft (DFG, German Research Foundation, project numbers 447948357 and 440958198) and the Sino-German Center for Research Promotion (Project M-0294).

\medskip

%

\textbf{References}\\

1	R. Horodecki, P. Horodecki, M. Horodecki, and K. Horodecki, Quantum entanglement, Rev. Mod. Phys. 81, 865 (2009).\\
2	M. A. Nielsen and I. L. Chuang, Quantum computation and quantum information (Cambridge University Press, Cambridge, U.K., 2000).\\
3	J.-W. Pan, C. Simon, C. Brukner, and A. Zeilinger, Entanglement purification for quantum communication, Nature 410, 1067 (2001).\\
4	R. Ursin, F. Tiefenbacher, T. Schmitt-Manderbach, H. Weier, T. Scheidl, M. Lindenthal, B. Blauensteiner, T. Jennewein, J. Perdigues, P. Trojek, B. \"{O}mer, M. F\"{u}rst, M. Meyenburg, J. Rarity, Z. Sodnik, C. Barbieri, H. Weinfurter and A. Zeilinger , Entanglement-based quantum communication over 144 km, Nat. Phys. 3, 481 (2007).\\
5	W. J. Munro, A. M. Stephens, S. J. Devitt, K. A. Harrison, and K. Nemoto, Quantum communication without the necessity of quantum memories, Nat. Photonics 6, 777 (2012).\\
6	C. Couteau, S. Barz, T. Durt, T. Gerrits, J. Huwer, R. Prevedel, J. Rarity, A. Shields, and G. Weihs, Applications of single photons to quantum communication and computing, Nat. Rev. Phys. 5, 326 (2023).\\
7	V. Scarani, H. Bechmann-Pasquinucci, N. J. Cerf, M. D\v{u}sek, N. L\"{u}tkenhaus, and M. Peev, The security of practical quantum key distribution, Rev. Mod. Phys. 81, 1301 (2009).\\
8	V. Zapatero, T. van Leent, R. Arnon-Friedman, W.-Z. Liu, Q. Zhang, H. Weinfurter, and M. Curty, Advances in device-independent quantum key distribution, npj Quantum Inf. 9 (2023).\\
9	Y.-Z. Zhen, Y. Mao, Y.-Z. Zhang, F. Xu, and B. C. Sanders, Device-independent quantum key distribution based on the Mermin-Peres magic square game, Phys. Rev. Lett. 131, 8 080801 (2023).\\
10	A. Karlsson and M. Bourennane, Quantum teleportation using three-particle entanglement, Phys. Rev. A 58, 4394 (1998).\\
11	X.-M. Hu, C. Zhang, B.-H. Liu, Y. Cai, X.-J. Ye, Y. Guo, W.-B. Xing, C.-X. Huang, Y.-F. Huang, C.-F. Li, and G.-C. Guo, Experimental high-dimensional quantum teleportation, Phys. Rev. Lett. 125, 230501 (2020).\\
12	A. Harrow, P. Hayden, and D. Leung, Superdense coding of quantum states, Phys. Rev. Lett. 92, 187901 (2004).\\
13	J. Tang, Q. Zeng, N. Feng, and Z. Wang, Superdense coding based on intraparticle entanglement states, Eur. Phys. J. D 76, 172 (2022).\\
14	A. M. Steane, Efficient fault-tolerant quantum computing, Nature 399, 124 (1999).\\
15	M. A. Nielsen and C. M. Dawson, Fault-tolerant quantum computation with cluster states, Phys. Rev. A 71, 042323 (2005).\\
16	K. Fukui, High-threshold fault-tolerant quantum computation with the Gottesman-Kitaev-Preskill qubit under noise in an optical setup, Phys. Rev. A 107, 052414 (2023).\\
17	S. Chakraborty, S. Banerjee, S. Adhikari, and A. Kumar, Entanglement in the Grover’s search algorithm, arXiv:1305.4454 (2013).\\
18	B. P. Lanyon, T. J. Weinhold, N. K. Langford, M. Barbieri, D. F. V. James, A. Gilchrist, and A. G. White, Experimental demonstration of a compiled version of Shor’s algorithm
with quantum entanglement, Phys. Rev. Lett. 99, 250505 (2007).\\
19	C. H. Bennett, D. P. DiVincenzo, J. A. Smolin, and W. K. Wootters, Mixed-state entanglement and quantum error correction, Phys. Rev. A 54, 3824 (1996).\\
20	A. Ekert and C. Macchiavello, Quantum error correction for communication, Phys. Rev. Lett. 77, 2585 (1996).\\
21	A. J. Scott, Multipartite entanglement, quantum-error-correcting codes, and entangling power of quantum evolutions, Phys. Rev. A 69, 052330 (2004).\\
22	C. Galindo, F. Hernando, and D. Ruano, Entanglement-assisted quantum error-correcting codes from RS codes and BCH codes with extension degree 2, Quantum Inf. Process. 20, 158 (2021).\\
23	P. Mazurek, M. Farkas, A. Grudka, M. Horodecki, and M. Studzi\'{n}ski, Quantum error-correction codes and absolutely maximally entangled states, Phys. Rev. A 101, 042305 (2020).\\
24	D. F. V. James, P. G. Kwiat, W. J. Munro, and A. G. White, Measurement of qubits, Phys. Rev. A 64, 052312 (2001).\\
25	A. I. Lvovsky and M. G. Raymer, Continuous-variable optical quantum-state tomography, Rev. Mod. Phys. 81, 299 (2009).\\
26	M. Rambach, M. Qaryan, M. Kewming, C. Ferrie, A. G. White, and J. Romero, Robust and efficient high-dimensional quantum state tomography, Phys. Rev. Lett. 126, 100402 (2021).\\
27	S. T. Flammia and Y.-K. Liu, Direct fidelity estimation from few Pauli measurements, Phys. Rev. Lett. 106, 230501 (2011).\\
28	M. P. da Silva, O. Landon-Cardinal, and D. Poulin, Practical characterization of quantum devices without tomography, Phys. Rev. Lett. 107, 210404 (2011).\\
29	S. Pallister, N. Linden, and A. Montanaro, Optimal verification of entangled states with local measurements, Phys. Rev. Lett. 120, 170502 (2018).\\
30	Z. Li, Y.-G. Han, and H. Zhu, Efficient verification of bipartite pure states, Phys. Rev. A 100, 032316 (2019).\\
31	Y.-C. Liu, X.-D. Yu, J. Shang, H. Zhu, and X. Zhang, Efficient verification of Dicke states, Phys. Rev. Appl. 12, 044020 (2019).\\
32	Z. Li, Y.-G. Han, H.-F. Sun, J. Shang, and H. Zhu, Verification of phased Dicke states, Phys. Rev. A 103, 022601 (2021).\\
33	H. Zhu and M. Hayashi, Efficient verification of hypergraph states, Phys. Rev. Appl. 12, 054047 (2019).\\
34	X.-D. Yu, J. Shang, and O. G\"{u}hne, Optimal verification of general bipartite pure states, npj Quantum Inf. 5, 112 (2019).\\
35	Z. Li, Y.-G. Han, and H. Zhu, Optimal verification of Greenberger-Horne-Zeilinger states, Phys. Rev. Appl. 13, 054002 (2020).\\
36	N. Dangniam, Y.-G. Han, and H. Zhu, Optimal verification of stabilizer states, Phys. Rev. Res. 2, 043323 (2020).\\
37	H. Zhu and M. Hayashi, Efficient verification of pure quantum states in the adversarial scenario, Phys. Rev. Lett. 123, 260504 (2019).\\
38	H. Zhu and M. Hayashi, General framework for verifying pure quantum states in the adversarial scenario, Phys. Rev. A 100, 062335 (2019).\\
39	Z. Li, H. Zhu, and M. Hayashi, Robust and efficient verification of graph states in blind measurement-based quantum computation, npj Quantum Inf. 9 (2023).\\
40	Y.-C. Liu, J. Shang, and X. Zhang, Efficient verification of entangled continuous-variable quantum states with local measurements, Phys. Rev. Res. 3, L042004 (2021).\\
41	Y.-C. Liu, J. Shang, R. Han, and X. Zhang, Universally optimal verification of entangled states with nondemolition measurements, Phys. Rev. Lett. 126, 090504 (2021).\\
42	H. Zhu and H. Zhang, Efficient verification of quantum gates with local operations, Phys. Rev. A 101, 042316 (2020).\\
43	Y.-C. Liu, J. Shang, X.-D. Yu, and X. Zhang, Efficient verification of quantum processes, Phys. Rev. A 101, 042315 (2020).\\
44	J. Preskill, Quantum Computing in the NISQ era and beyond, Quantum 2, 79 (2018).\\
45	W.-H. Zhang, C. Zhang, Z. Chen, X.-X. Peng, X.-Y. Xu, P. Yin, S. Yu, X.-J. Ye, Y.-J. Han, J.-S. Xu, G. Chen, C.-F. Li, and G.-C. Guo, Experimental optimal verification of entangled states using local measurements, Phys. Rev. Lett. 125, 030506 (2020).\\
46	X. Jiang, K. Wang, K. Qian, Z. Chen, Z. Chen, L. Lu, L. Xia, F. Song, S. Zhu, and X. Ma, Towards the standardization of quantum state verification using optimal strategies, npj Quantum Inf. 6, 90 (2020).\\
47	W.-H. Zhang, X. Liu, P. Yin, X.-X. Peng, G.-C. Li, X.-Y. Xu, S. Yu, Z.-B. Hou, Y.-J. Han, J.-S. Xu, et al., Classical communication enhanced quantum state verification, npj Quantum Inf. 6, 103 (2020).\\
48	L. P. Thinh, M. Dall’Arno, and V. Scarani, Worst-case quantum hypothesis testing with separable measurements, Quantum 4, 320 (2020).\\
49	F. B. Maciejewski, Z. Zimbor\'{a}s, and M. Oszmaniec, Mitigation of readout noise in near-term quantum devices by classical post-processing based on detector tomography, Quantum 4, 257 (2020).


\appendix

\section{Proof of Observation 1}\label{app:UniNoise}
Here we prove that the distinguishable conditions are necessary and sufficient for a strategy $\Omega$ to uniquely identify the target state $\ket{\psi}$ from any other state.

\begin{proof}
Suppose the spectral decomposition of a strategy $\Omega$ is
\begin{equation}
    \Omega=\sum_{i=0}^{d-1}{\lambda_i}\ket{\phi_i}\bra{\phi_i}\,,
\end{equation}
where $\lambda_i\ge\lambda_j$ for $i<j$ and $i \neq 0$. $\big\{\ket{\phi_i}\big\}$ form an orthonormal basis in the Hilbert space. 

Firstly, we prove the sufficiency of the distinguishable conditions, i.e., that if the strategy satisfies the distinguishable conditions, where the target state $\ket{\psi}$ is the dominant eigenstate $\ket{\phi_0}$ with eigenvalue $\lambda_0$, then it can uniquely distinguish the target state from any other state.
For any pure state $\ket{\phi^{\prime}}$ other than $\ket{\phi_0}$, its decomposition reads
\begin{equation}
\ket{\phi^{\prime}}=a\ket{\phi_0}+b\ket{\phi_0^{\bot}}
\end{equation}
where $\ket{\phi_0^{\bot}}$ is orthogonal to $\ket{\phi_0}$, $\left | a \right |^2 + \left | b \right |^2 = 1$, $|b|>0$. Direct calculation shows
\begin{equation}
\begin{aligned}
\tr\bigl(\Omega\ket{\phi^{\prime}}\bra{\phi^{\prime}}\bigr)&=\bra{\phi^{\prime}}\Omega\ket{\phi^{\prime}}\\
&=a^{*}a\bra{\phi_0}\Omega \ket{\phi_0}+a^{*}b\bra{\phi_0}\Omega \ket{\phi_0^{\bot}}+b^{*}a\bra{\phi_0^{\bot}}\Omega \ket{\phi_0}+b^{*}b\bra{\phi_0^{\bot}}\Omega \ket{\phi_0^{\bot}}\\
&=\left | a \right |^2\lambda_0+\left | b \right |^2\bra{\phi_0^{\bot}}\Omega \ket{\phi_0^{\bot}}<\lambda_0\,.
\end{aligned}
\end{equation}
So for any pure state $\ket{\phi^{\prime}}$ other than $\ket{\phi_0}$, the probability $\tr(\Omega\ket{\phi^{\prime}}\bra{\phi^{\prime}})$ must be less than $\lambda_0$.
Moreover, for any mixed state $\rho=\sum_ip_i\ket{\psi_i}\bra{\psi_i}$,
\begin{equation}
\tr(\rho\Omega)=\sum_ip_i\bigl(\Omega\ket{\psi_i}\bra{\psi_i}\bigr)
<\sum_ip_i\bigl(\Omega\ket{\phi_0}\bra{\phi_0}\bigr)=\lambda_0\,.
\end{equation}
Thus, strategy $\Omega$ can distinguish any other state from the target state $\ket{\psi}$.

Next, we show that the distinguishable conditions are necessary for a noisy strategy $\Omega$ to uniquely identify the target state $\ket{\psi}$. 
If the dominant eigenstate $\ket{\phi_0}$ is not the target state $\ket{\psi}$, then  
\begin{equation}
    \ket{\psi} = \sqrt{1-\varepsilon}\ket{\phi_0} + \sqrt{\varepsilon}\ket{\phi_0^{\bot}}\,,
\end{equation}
where $0<\varepsilon\le 1$.
There exists another state
\begin{equation}
    \ket{\psi'} = \sqrt{1-\varepsilon}\ket{\phi_0} - \sqrt{\varepsilon}\ket{\phi_0^{\bot}}\,,
\end{equation}
such that 
\begin{equation}\label{app:eqA7}
    \tr\bigl(\Omega\ket{\psi'}\bra{\psi'}\bigr) = \tr\bigl(\Omega\ket{\psi}\bra{\psi}\bigr)\,.
\end{equation}
If the spectral gap equals to 0 and the target state $\ket{\psi}$ is an eigenstate of $\Omega$ with the largest eigenvalue $\lambda_0$, any $\ket{\psi'}$ in the degenerate eigensubspace for $\lambda_0$ satisfies \textbf{Equation 36}.
Thus, a noisy verification strategy $\tilde{\Omega}$ can uniquely distinguish the target state \emph{iff} it meets the distinguishable conditions.
\end{proof}

\section{Proof of Equation 3 and 4}\label{App:B}
Here we prove that if the distinguishable conditions are met, the worst-case state is in the form of a pure state as in Equation 3.

\begin{proof}
Under the distinguishable conditions, the spectral decomposition of a noisy strategy $\tilde{\Omega}$ could be written as
\begin{equation}
\tilde{\Omega}=\lambda_0\ket{ \psi}\bra{ \psi}+\sum_{i=1}^{d-1}\lambda_i\ket{ \psi^{\perp}_{i} }\bra{ \psi^{\perp}_{i}}\,,
\end{equation}
where $\ket{ \psi}$ is the target state with dominant eigenvalue $\lambda_0$, and the worst-case state $\sigma$ reads
\begin{equation}
    \sigma=r\ket{ \psi}\bra{ \psi}+(1-r)\sigma^{\perp}+c\ket{ \psi}\bra{\Phi^{\perp}}+c^{*}\ket{\Phi^{\perp}}\bra{ \psi}\,.
\end{equation}
$\ket{\Phi^\perp}$ is a state in the subspace spanned by $\ket{\psi^{\perp}_{i}}$. The worst-case $\sigma$ maximizes $\tr(\tilde{\Omega}\sigma)$ as follows
\begin{equation}\label{optiso}
\max_{\sigma}\tr(\tilde{\Omega}\sigma)=\max_{\sigma}\Bigg[\lambda_0\bra{ \psi}\sigma\ket{ \psi}+\tr\biggl(\sum_i\lambda_i\ket{ \psi^{\perp}_{i} }\bra{ \psi^{\perp}_{i} }\sigma\biggr)\Bigg]
    =r\lambda_0+(1-r)\lambda_1
    =\lambda_1+r\bigr(\lambda_0-\lambda_{1}\bigl)\,,
\end{equation}
where the optimal $\sigma^{\perp}=\ket{ \psi^{\perp}_{1}}\bra{ \psi^{\perp}_{1}}$ is the eigenstate corresponding to the second-largest eigenvalue $\lambda_{1}$ of  $\tilde{\Omega}$. Note $\bra{ \psi}\sigma\ket{ \psi}\leq1-\epsilon$, implying $r\leq1-\epsilon$. Therefore, $\tr(\tilde{\Omega}\sigma)=\lambda_{1}+(1-\epsilon)(\lambda_0-\lambda_{1})=\lambda_0-\epsilon(\lambda_0-\lambda_{1})$ is maximized by choosing $r=1-\epsilon$. Hence, the worst-case scenario could be achieved by a pure state $\sigma=\ket{ \psi^{\prime}}\bra{ \psi^{\prime}}$, where $\ket{ \psi^{\prime}}=\sqrt{1-\epsilon}\ket{ \psi}+\sqrt{\epsilon}\ket{ \psi^{\perp}_{1}}$. Be noted that $\sigma$ is not uniquely the pure state $\psi^{\prime}$ because the states corresponding to the optimal solution of \textbf{Equation 39} are not unique. However, Equation 39 shows that all possible states correspond to the same probability, which is $\lambda_1 + r(\lambda_0 - \lambda_1)$. Therefore, the pure state $\ket{ \psi^{\prime}} = \sqrt{1-\epsilon}\ket{ \psi} + \sqrt{\epsilon}\ket{ \psi^{\perp}_{1}}$ can be regarded as an example of all possible worst-case states $\sigma$, but not as a unique solution.
\end{proof}

\section{Discussion of Chernoff bound}\label{Chernoff}
To establish an explicit relation between the sample complexity and noise, we employ the Chernoff bound to characterize the binomial cumulative distribution, i.e.,
\begin{equation}\label{eq:chboundfn}
    F^{\leftarrow}(k ; N, p)=\operatorname{Pr}(X \leq k)
    \le e^{-\cD\bigl(\,\frac{k}{N}\mid\mid p\,\bigr)N}\,,
\end{equation}
where
\begin{equation}
\cD(f \mid\mid p):=f \ln \frac{f}{p}+(1-f) \ln \frac{1-f}{1-p}
\end{equation}
is the Kullback–Leibler divergence. Thus, the average error rate $\rP_\text{sym}$ in \textbf{Equation 8} can be bounded as
\begin{equation}
\begin{aligned}
    \rP_\text{sym}&\leq\frac{e^{-\cD\bigl(\,\lambda_0\mid\mid\lambda_0-\frac{1}{2}\nu(\tilde{\Omega})\epsilon\,\bigr)N}+e^{-\cD\bigl(\,\lambda_0-\nu(\tilde{\Omega})\epsilon\mid\mid\lambda_0-\frac{1}{2}\nu(\tilde{\Omega})\epsilon\,\bigr)N}}{2}\\
    &=\frac{1}{2}\left[\biggl(\frac{\lambda_0-\frac{1}{2}\nu(\tilde{\Omega})\epsilon}{\lambda_0}\biggr)^{\lambda_0}\biggl(\frac{1-\lambda_0+\frac{1}{2}\nu(\tilde{\Omega})\epsilon}{1-\lambda_0}\biggr)^{1-\lambda_0}\right]^N
\\
&\phantom{=} +\frac{1}{2}\left[\biggl(\frac{\lambda_0-\frac{1}{2}\nu(\tilde{\Omega})\epsilon}{\lambda_0-\nu(\tilde{\Omega})\epsilon}\biggr)^{\lambda_0-\nu(\tilde{\Omega})\epsilon}\biggl(\frac{1-\lambda_0+\frac{1}{2}\nu(\tilde{\Omega})\epsilon}{1-\lambda_0+\nu(\tilde{\Omega})\epsilon}\biggr)^{1-\lambda_0+\nu(\tilde{\Omega})\epsilon}\right]^N\,.
\end{aligned} 
\end{equation}
In comparison to the standard QSV efficiency $(1-\nu(\tilde{\Omega})\epsilon)^N$, we could examine $\sqrt[N]{\rP_\text{sym}}$ first:
\begin{equation}
\begin{aligned}
\sqrt[N]{\rP_\text{sym}}&\leq\left\{\frac{1}{2}\left[\biggl(\frac{\lambda_0-\frac{1}{2}\nu(\tilde{\Omega})\epsilon}{\lambda_0}\biggr)^{\lambda_0}\biggl(\frac{1-\lambda_0+\frac{1}{2}\nu(\tilde{\Omega})\epsilon}{1-\lambda_0}\biggr)^{1-\lambda_0}\right]^N\right.\\
& \phantom{=}
+\left.\frac{1}{2}\left[\biggl(\frac{\lambda_0-\frac{1}{2}\nu(\tilde{\Omega})\epsilon}{\lambda_0-\nu(\tilde{\Omega})\epsilon}\biggr)^{\lambda_0-\nu(\tilde{\Omega})\epsilon}\biggl(\frac{1-\lambda_0+\frac{1}{2}\nu(\tilde{\Omega})\epsilon}{1-\lambda_0+\nu(\tilde{\Omega})\epsilon}\biggr)^{1-\lambda_0+\nu(\tilde{\Omega})\epsilon}\right]^N\right\}^{\frac{1}{N}}\\
&=1+0*\nu(\tilde{\Omega})\epsilon-\frac{1}{8\bigl(1-\lambda_0\bigr)\lambda_0}*\bigl(\nu(\tilde{\Omega})\epsilon\bigr)^2+...+R_n\bigl(\nu(\tilde{\Omega})\epsilon\bigr)\,.
\end{aligned}
\end{equation}
Thus,
\begin{equation}
\rP_\text{sym}={\biggl(\sqrt[N]{\rP_\text{sym}}\biggr)}^N
    \leq\bigg[1+0*\nu(\tilde{\Omega})\epsilon-\frac{1}{8\bigl(1-\lambda_0\bigr)\lambda_0}*(\nu(\tilde{\Omega})\epsilon)^2+...+R_n(\nu(\tilde{\Omega})\epsilon)\bigg]^N\,.
\end{equation}
If we ignore high-order items and require $\bigg[1-\frac{1}{8\bigl(1-\lambda_0\bigr)\lambda_0}*(\nu(\tilde{\Omega})\epsilon)^2\bigg]^N\le\delta$, then
\begin{equation}\label{eq10}
    N \geq \frac{\ln \delta}{\ln \bigg(1-\frac{1}{8\bigl(1-\lambda_0\bigr)\lambda_0}*(\nu(\tilde{\Omega})\epsilon)^2\bigg)} \approx 8\bigl(1-\lambda_0\bigr)\lambda_0(\nu(\tilde{\Omega})\epsilon)^{-2} \ln \delta^{-1}\,
\end{equation}
shows that the sample complexity $N$ exhibits a negative quadratic relation with both the infidelity $\epsilon$ and the spectral gap $\nu(\tilde{\Omega})$ of the noisy strategy $\tilde{\Omega}$. This implies that the sample complexity of noisy QSV remains polynomial with respect to the size of the quantum system. Additionally, apart from the spectral gap $\nu(\tilde{\Omega})$ and infidelity $\epsilon$, measurement noise further influences the sample complexity by affecting the dominant eigenvalue $\lambda_0$.

\section{Proof of Proposition 1}\label{Stabilizer}
Here we prove that when the assumptions in Equation 20 and 24 hold true, the strategy $\Omega_{\mathrm{S}}$ in Equation 19 satisfies the dintinguishable conditions.

\begin{proof}
Recall that the stabilizer state $\ket{\psi}$ is defined by its stabilizer group $\{G_k\}$, the generators of which are denoted by $\langle S_j\rangle$.
The readout noise under the assumptions in Equation 20 and 24 only introduces a noise factor $g_k=\prod_i^n(1-2\eta_i)$ on the operator $G_k$, as stated in \textbf{Equation 25}. Then for any test operator $E_k=\frac{1}{2}\bigl(G_k+\mathbbm{1}\bigr)$, its noisy version $\tilde{E}_k$ reads
\begin{equation}
\tilde{E}_k=\frac{1}{2}\Bigl(\tilde{G}_k+\mathbbm{1}\Bigr)
    =\frac{1}{2}\bigl(g_kG_k+\mathbbm{1}\bigr)\,.
\end{equation}
The noisy measurement operator $\tilde{E}_k$ has only two eigenvalues: $\frac{1}{2}\bigl(G_k+1\bigr)$ and $\frac{1}{2}\bigl(1-G_k\bigr)$, where $G_k$ is the noise parameter. The stabilizer state $\ket{\psi}$ is the eigenstate with the greater eigenvalue $\frac{1}{2}\bigl(G_k+1\bigr)$. Moreover, the stabilizer state is the unique state that has the larger eigenvalue for each test operator $\tilde{E}_k$, due to the fact that the dimension of the common eigenspace of the stabilizer group elements is $d_\mathcal{H_G}=1$. Thus, the dominant eigenvalue of the noisy strategy $\tilde{\Omega}_{\mathrm{S}}$ is $\frac{1}{2^n-1}\sum_k\frac{1}{2}(G_k+1)$. Note that $\big\{E_k\big\}$ is a commutative set, consequently, $\Big\{\tilde{E}_k\Big\}$ are also commutative and share the same common eigenspace $\bigl\{\ket{G_{\mathbf{w}}}\bigr\}$. $\bigl\{\ket{G_{\mathbf{w}}}\bigr\}$ form the stabilizer basis labeled by vectors in the binary vector space $\mathbf{w}\in\mathbbm{Z}_2^{n}$, as follows
\begin{equation}
\ket{G_{\mathbf{w}}}\bra{G_{\mathbf{w}}}=\Pi_{G_{\mathbf{w}}}=\prod_{j=1}^{n} \frac{\mathbbm{1}+(-1)^{w_{j}} S_{j}}{2}=\sum_{\mathbf{y}\in \mathbbm{Z}_{2}^{n}}(-1)^{\mathbf{w}\cdot \mathbf{y}} S^{\mathbf{y}}\,,
\end{equation}
where $S^{\mathbf{y}}=\prod_{j=1}^{n}S_j^{y_j}$ is an element in group $\big\{G_k\big\}$ denoted by $\mathbf{y}\in\mathbbm{Z}_2^{n}$. Thus, for each stabilizer basis $\ket{G_{\mathbf{w}}}$, its pass probability of the measurement $\tilde{E}_k=\frac{g_kG_k+\mathbbm{1}}{2}$ reads
\begin{equation}
\begin{aligned}
\tr\biggl(\tilde{E}_k\ket{G_{\mathbf{w}}}\bra{G_{\mathbf{w}}}\biggr)&=\tr\biggl(\sum_{\mathbf{y} \in \mathbbm{Z}_{2}^{n}}(-1)^{\mathbf{w}\cdot \mathbf{y}} S^{\mathbf{y}}\tilde{E}_k\biggr)=\frac{1}{2}\tr\Bigg[\sum_{\mathbf{y} \in \mathbbm{Z}_{2}^{n}}(-1)^{\mathbf{w}\cdot \mathbf{y}} S^{\mathbf{y}}(g_kG_k+\mathbbm{1})\Bigg]\\
&=\frac{1}{2}+\frac{g_k}{2}\tr\biggl(\sum_{\mathbf{y} \in \mathbbm{Z}_{2}^{n}}(-1)^{\mathbf{w}\cdot \mathbf{y}} S^{\mathbf{y}}S^{\mathbf{k}}\biggr)=\frac{1}{2}\biggl(1+g_k(-1)^{\mathbf{w}\cdot \mathbf{k}}\biggr)\,.
\end{aligned}
\end{equation}
Here, $G_k=S^{\mathbf{k}}$ denotes an element in the group $\big\{G_k\big\}$, excluding the identity element $\mathbbm{1}$, where $\mathbf{k}\in\mathbbm{Z}_{2}^{n}\setminus\{\mathbf{0}\}$. Thus, the probability $p_{\mathbf{w}}$ of passing through the noisy strategy $\tilde{\Omega}_{\mathrm{S}}$ for $\ket{G_{\mathbf{w}}}$ is
\begin{equation}
p_{\mathbf{w}}=\tr\biggl(\tilde{\Omega}_{\mathrm{S}}\ket{G_{\mathbf{w}}}\bra{G_{\mathbf{w}}}\biggr)
=\frac{1}{2^n-1}\sum_{\mathbf{k}}\frac{1}{2}\Bigl(1+g_k(-1)^{\mathbf{w}\cdot \mathbf{k}}\Bigr)\,.
\end{equation}
The maximum eigenvalue $\lambda_{1}$ can be obtained
\begin{equation}
\lambda_{1}=\max_{\mathbf{w}\in\mathbbm{Z}_{2}^{n}\setminus\{\mathbf{0}\}}p_{\mathbf{w}}\,.
\end{equation}

\end{proof}

\section{Proof of Proposition 2}\label{GHZ}
Here we prove that when the assumptions in Equation 20,  24, and 53 are met, the strategy $\Omega_{\mathrm{I}}$ in \textbf{Equation 27} could satisfy the dintinguishable conditions.

\begin{proof}
We replace all noiseless local Pauli measurements, denoted as $\{\Pi_{X+}, \Pi_{X-}, \Pi_{Y+}, \Pi_{Y-}, \Pi_{Z+}, \Pi_{Z-}\}$, with their respective noisy version represented by $\{\tilde{\Pi}_{X+}, \tilde{\Pi}_{X-}, \tilde{\Pi}_{Y+}, \tilde{\Pi}_{Y-}, \tilde{\Pi}_{Z+}, \tilde{\Pi}_{Z-}\}$. For instance, noisy measurements of Pauli-$X$ could be written as
\begin{equation}
\begin{aligned}
\left(\begin{array}{c}
\tilde{\Pi}_{X+} \\
\tilde{\Pi}_{X-}
\end{array}\right)&=\Lambda_{x}\left(\begin{array}{l}
\Pi_{X+} \\
\Pi_{X-}
\end{array}\right) =\left(\begin{array}{cc}
1-\eta_x & \eta_x \\
\eta_x & 1-\eta_x
\end{array}\right)
\left(\begin{array}{l}
\Pi_{X+} \\
\Pi_{X-}
\end{array}\right)\,,
\end{aligned}
\end{equation}
where $\eta_x$ is the noisy parameter of Pauli-$X$ measurement.

For test $P_{0}$ in the verification operator $\Omega_{\mathrm{I}}$, the noisy version is
\begin{equation}\label{A2}
\tilde{P}_0     ={\otimes}_i\bigl(\tilde{\Pi}_{Z+}\bigr)_i+{\otimes}_i\bigl(\tilde{\Pi}_{Z-}\bigr)_i
     ={\otimes}_i\left[\Bigl(1-\eta^i_z\Bigr)\Pi^i_{Z+}+\eta^i_z\Pi^i_{Z-}\right]
     +{\otimes}_i\left[\eta^i_z\Pi^i_{Z+}+\Bigl(1-\eta^i_z\Bigr)\Pi^i_{Z-}\right]\,.
\end{equation}
The $\eta^i_z$ is noisy parameter of measurement on $i$-th qubit and $\bigl(\tilde{\Pi}_{Z+}\bigr)_i$ means noisy projective measurement on $i$-th qubit. 

For the first item ${\otimes}_i\left[\Bigl(1-\eta^i_z\Bigr)\Pi^i_{Z+}+\eta^i_z\Pi^i_{Z-}\right]$ in Equation 52.
It is straightforward to verify that the state $\ket{0}^{\otimes n}$ has an eigenvalue of $\prod_{i=1}^{n} \Bigl(1 - \eta^i_z\Bigr)$, while the state $\ket{1}^{\otimes n}$ has an eigenvalue of $\prod_{i=1}^{n} \eta^i_z$. Similarly, for the second term in Equation 52, the state $\ket{1}^{\otimes n}$ has an eigenvalue of $\prod_{i=1}^{n} \Bigl(1 - \eta^i_z\Bigr)$, and the state $\ket{0}^{\otimes n}$ has an eigenvalue of $\prod_{i=1}^{n} \eta^i_z$. Thus, the GHZ state is an eigenspace of the noisy operator $\tilde{P}_0$ with eigenvalue $\prod^{n}_{i=1}\Bigl(1-\eta^i_z\Bigr)+\prod^{n}_{i=1}\eta^i_z$. We will then demonstrate the condition under which $\prod^{n}_{i}\Bigl(1-\eta^i_z\Bigr)+\prod^{n}_{i}\eta^i_z$ becomes the largest eigenvalue of the test $\tilde{P}_0$.

Note that the eigenvalue of $\tilde{P}_0$ could be expressed as $\prod\limits_{i\in A}\Bigl(1-\eta^i_z\Bigr)\prod\limits_{i\in\overline{A}}\eta^i_z+\prod\limits_{i\in \overline{A}}\Bigl(1-\eta^i_z\Bigr)\prod\limits_{i\in A}\eta^i_z$, where $A\in\{1,2,\cdots,n\}$ represents a combination of qubits from the set $\{1,2,\cdots,n\}$. When the first item ${\otimes}_i\left[\Bigl(1-\eta^i_z\Bigr)\Pi^i_{Z+}+\eta^i_z\Pi^i_{Z-}\right]$ in Equation 52 takes one value of the $i$-th qubit $\Big\{\eta^i_z,1-\eta^i_z\Big\}$, the second item ${\otimes}_i\left[\eta^i_z\Pi^i_{Z+}+\Bigl(1-\eta^i_z\Bigr)\Pi^i_{Z-}\right]$ can only take the other. If we require that $\prod^{n}_{i=1}\Bigl(1-\eta^i_z\Bigr)+\prod^{n}_{i=1}\eta^i_z$ is the uniquely largest eigenvalue among all possible combinations $\prod\limits_{i\in A}\Bigl(1-\eta^i_z\Bigr)\prod\limits_{i\in\overline{A}}\eta^i_z+\prod\limits_{i\in \overline{A}}\Bigl(1-\eta^i_z\Bigr)\prod\limits_{i\in A}\eta^i_z$ depending on the set $A$, the following condition should be satisfied for all sets $A$ that do not correspond to the largest eigenvalue $\prod^{n}_{i=1}\Bigl(1-\eta^i_z\Bigr)+\prod^{n}_{i=1}\eta^i_z$:

\begin{equation}\label{GHZGEQ}
	\begin{aligned}
		&\prod^{n}_{i=1}\Bigl(1-\eta^i_z\Bigr)+\prod^{n}_{i=1}\eta^i_z-
		\Bigg[\prod\limits_{i\in A}\Bigl(1-\eta^i_z\Bigr)\prod\limits_{i\in\overline{A}}\eta^i_z+\prod\limits_{i\in \overline{A}}\Bigl(1-\eta^i_z\Bigr)\prod\limits_{i\in A}\eta^i_z\Bigg]\\
		=&\Bigg[\prod^{n}_{i=1}\Bigl(1-\eta^i_z\Bigr)-\prod\limits_{i\in A}\Bigl(1-\eta^i_z\Bigr)\prod\limits_{i\in\overline{A}}\eta^i_z\Bigg]
		+\Bigg[\prod^{n}_{i=1}\eta^i_z-\prod\limits_{i\in \overline{A}}\Bigl(1-\eta^i_z\Bigr)\prod\limits_{i\in A}\eta^i_z\Bigg]\\
		=&\Bigg[\prod\limits_{i\in A}\Bigl(1-\eta^i_z\Bigr)\Biggl(\prod\limits_{i\in\overline{A}}\Bigl(1-\eta^i_z\Bigr)-\prod\limits_{i\in\overline{A}}\eta^i_z\Biggr)\Bigg]-\Bigg[\Biggl(\prod\limits_{i\in\overline{A}}\Bigl(1-\eta^i_z\Bigr)-\prod\limits_{i\in\overline{A}}\eta^i_z\Biggr)\prod\limits_{i\in A}\eta^i_z\Bigg]\\
		=&\Biggl(\prod\limits_{i\in\overline{A}}\Bigl(1-\eta^i_z\Bigr)-\prod\limits_{i\in\overline{A}}\eta^i_z\Biggr)\Biggl(\prod\limits_{i\in A}\Bigl(1-\eta^i_z\Bigr)-\prod\limits_{i\in A}\eta^i_z \Biggr)> 0 \,.
	\end{aligned}
\end{equation}
\textbf{Equation 53} guarantees that the eigenvalue of the GHZ state $\prod^{n}_{i=1}\Bigl(1-\eta^i_z\Bigr)+\prod^{n}_{i=1}\eta^i_z$ is the uniquely largest eigenvalue among all the eigenvalues of noisy measurement $\tilde{P}_0$.
The test $P_{\mathcal{Y}}$ in the strategy $\Omega_{\mathrm{I}}$ is similar to the test in the stabilizer state verification strategy $\Omega_{\mathrm{S}}$. Thus, Proposition 2 establishes that the GHZ state takes the dominant eigenvalue $\frac{1}{3}\left[\prod_{i=1}^{n}\Bigl(1-\eta^i_z\Bigr)+\prod_{i=1}^{n}\eta^i_z\right]+\frac{1}{3\cdot 2^{n-2}}\sum_{\mathcal{Y}} \frac{1}{2}(g_{\mathcal{Y}}+1)$ of the noisy strategy $\tilde{\Omega}_{\mathrm{I}}$, where $g_{\mathcal{Y}}$ denotes the noisy parameter dependent on the test $P_{\mathcal{Y}}$.

\end{proof}

\end{document}